\documentclass{article}
\usepackage[T1]{fontenc}
\usepackage[latin1]{inputenc}
\usepackage{geometry}
\usepackage{setspace}
\makeatletter

\providecommand{\LyX}{L\kern-.1667em\lower.25em\hbox{Y}\kern-.125emX\@}
\newcommand{\noun}[1]{\textsc{#1}}

 \newcommand{\lyxaddress}[1]{
   \par {\raggedright #1 
   \vspace{1.4em}
   \noindent\par}
 }

\makeatother

\begin{document}

\title{Some Observations on Non-covariant Gauges and the \( \epsilon  \)-term}

\author{Satish D. Joglekar}

\maketitle

\lyxaddress{Department of Physics, I.I.T.Kanpur, Kanpur 208016 {[}INDIA{]}}

\begin{abstract}
We consider the Lagrangian path-integrals in Minkowski space for gauges with
a residual gauge-invariance. From rather elementary considerations, we demonstrate
the necessity of inclusion of an \( \epsilon  \)-term (even) in the \emph{formal
treatments}, without which one may reach incorrect conclusions. We show, further,
that the \( \epsilon  \)-term \emph{can} contribute to the BRST WT-identities
in a nontrivial way (even as \( \epsilon \rightarrow 0 \)) . We also show that
the (expectation value of the) correct \( \epsilon  \)-term satisfies an algebraic
condition. We show by considering (a commonly used) example of a simple local
quadratic \( \epsilon  \)-term, that they lead to additional constraints on
Green's function that are \emph{not} normally taken into account in the BRST
formalism that ignores the \( \epsilon  \)-term, and that they are characteristic
of the way the singularities in propagators are handled. We argue  that for
a subclass of these gauges, the Minkowski path-integral could not be obtained
by a Wick rotation from a Euclidean path-integral. 
\end{abstract}
The Yang-Mills theory in gauges other than the Lorentz gauges have been a subject
of wide research \cite{bns, l, 1}. These gauges have been used in a variety
of Standard Model calculations \cite{bns,l} . As compared to the covariant
gauges, these gauges have however not been fully developed. Recently, an approach
that gives the \emph{definition} of noncovariant gauges in a Lagrangian path-integral
formulation, that is compatible with the Lorentz gauges, has been given \cite{jm2000}.
It appears that several new observations regarding these gauges can be made
from simple and direct considerations suggested by hindsight received from these
works. This work also presents a general framework for dealing with these gauges
that may be of value.

\textbf{Notations:}

The Minkowski space Lagrangian path-integral for the Yang-Mills theory, (with
matter multiplets generically denoted by \( \psi  \)), and with a semi-simple
gauge group (with antisymmetric structure constants \( f^{\alpha \beta \gamma } \))
is usually formulated in terms of the generating functional as \cite{iz}\begin{equation}
\label{i}
W[J,K,\overline{K},\xi ,\overline{\xi }]=\int D\phi exp\{iS_{eff}[A,c,\overline{c},\psi ]+source-terms\}
\end{equation}
with the Faddeev-Popov effective action {[}FPEA{]} \begin{equation}
\label{eff}
S_{eff}=S_{0}+S_{gf}+S_{gh}
\end{equation}
where the gauge fixing term \( S_{gf} \) and the ghost term \( S_{gh} \) for
a gauge functional \( F^{\alpha } \) are given by\begin{equation}
\label{gf}
S_{gf}=\frac{-1}{2\lambda }\int d^{4}x\sum _{\alpha }F^{\alpha ^{2}}
\end{equation}
 \begin{eqnarray}
S_{gh}=-\int d^{4}xd^{4}y\, \overline{c}^{\alpha }(x)\frac{\delta F^{\alpha }(x)}{\delta A_{\mu }^{\beta }(y)}D_{\mu }^{\beta \gamma }c^{\gamma }(y) &  & \nonumber \\
\equiv -\int d^{4}xd^{4}y\, \overline{c}^{\alpha }(x)M^{\alpha \beta }(x,y;A)c^{\beta }(y) & \label{gh} 
\end{eqnarray}
and the source term reads\footnote{%
Below, for simplicity, we shall use symbols that do not distinguish between
fermion and scalar multiplets.
}\begin{equation}
\label{src}
i\int d^{4}x[JA+K^{\dagger }\psi +\psi ^{\dagger }K+\overline{\xi }c+\overline{c}\xi ]
\end{equation}
Further, the covariant derivative is defined as\[
D_{\mu }^{\alpha \beta }=\delta ^{\alpha \beta }\partial _{\mu }+gf^{\alpha \beta \gamma }A_{\mu }^{\gamma }\]
We may, from time to time, find it necessary to use the summation-integration
convention.

In addition to their use for covariant gauges, the path-integrals such as (\ref{i})
have widely been used for non-covariant gauges also. More importantly, they
have been employed in formal arguments in a variety of these gauges and contexts
\cite{bns,l}: (i) Axial gauges in Chern-Simon theory (ii) Coulomb gauges in
confinement problems \cite{egbz}, among several others. An essential property
of the FPEA, viz. the BRS invariance, leads via (\ref{i}) to the formal WT-identities.
These have extensively been used in various formulations in a \emph{formal}
manner, in formal arguments and the proof of unitarity and renormalizability
of various gauges \cite{bns,l} has also been attempted using these. String
theories also make use of such path-integral formulations\cite{gsw}.

In this work, we aim to bring out several non-trivial points in the context
of a wide class of gauges using only the general framework developed in earlier
works\cite{jm2000,jm}. They concern the necessity of keeping an \( \epsilon  \)-term
in the path-integrals and during its formal manipulations, importance of the
correct \( \epsilon  \)-term in the definition of the path-integral, \emph{non-trivial
contribution} of these \( \epsilon  \)-terms to the BRST \emph{WT-identities,}
additional constraints obeyed by the Green's functions in these gauges and the
possibility of Wick rotation in such path-integrals.

A number of simple results, labeled as the {}``theorems I-IV{}'' below, as
also a remark on Wick rotation hold for the {}``Type-R gauges{}'' to be defined
precisely below. These include the non-covariant gauges which have a residual
gauge transformation such as the Coulomb, the axial, the light-cone and the
temporal gauges. In this work, we shall present these results briefly; leaving
a detailed treatment to a future work \cite{prep}. 

\textbf{\noun{Definition}}: A gauge with a gauge functional \( F^{\alpha } \)
is called a \emph{{}``Type-R gauge{}''} if the following set of three conditions
are satisfied:

(i)There exists a set of infinitesimal gauge transformations characterized by
gauge parameters \( \sigma \theta ^{\alpha }(x) \), with \( \sigma  \) a constant
infinitesimal\footnote{%
Below, we will not use a notation that distinguishes between fermion and scalar
multiplets: the transformation laws can be rewritten in each context. Here,
\( T^{\alpha } \) are the Hermitian generators for an appropriate representation
of the gauge group.
} :\begin{equation}
\label{irgt}
\delta A_{\mu }^{\alpha }(x)=\sigma D_{\mu }^{\alpha \beta }\theta ^{\beta }(x);\delta \psi =ig\sigma T^{\alpha }\psi \theta ^{\alpha };\delta \psi ^{\dagger }=-ig\sigma \psi ^{\dagger }T^{\alpha }\theta ^{\alpha }
\end{equation}
such that \( S_{gf} \) of (\ref{gf}) is left invariant under them.

(ii) \( S_{gh} \) of (\ref{gh}) is invariant under the above gauge transformations
when combined with {}`` local vector{}'' transformations\footnote{%
These transformations have the form of the infinitesimal \emph{global} transformations;
but with an \emph{x-dependent} \( \theta ^{\alpha } \).
} on \( c \) and \( \overline{c} \) ; i.e. under\begin{eqnarray}
\delta A_{\mu }^{\alpha }(x)=\sigma D^{\alpha \beta }\theta ^{\beta }(x);\delta c^{\alpha }(x)=-\sigma gf^{\alpha \beta \gamma }c^{\beta }(x)\theta ^{\gamma }(x); &  & \nonumber \\
\delta \overline{c}^{\alpha }(x)=-\sigma gf^{\alpha \beta \gamma }\overline{c}^{\beta }(x)\theta ^{\gamma }(x) &  & \nonumber \\
\delta \psi =ig\sigma T^{\alpha }\psi \theta ^{\alpha };\delta \psi ^{\dagger }=-ig\sigma \psi ^{\dagger }T^{\alpha }\theta ^{\alpha } & \label{brs} 
\end{eqnarray}
In other words, the Faddeev-Popov effective term in the net action viz. \( \{-ilndetM\} \)
is invariant under the gauge transformations in (\ref{irgt}).

(iii) Under (\ref{brs}), the boundary conditions on the path-integral variables
in (\ref{i}) are unaltered as \( t\rightarrow \pm \infty  \) for all \( \mathbf{x} \).

We shall denote such a set as in (\ref{brs}), the infinitesimal residual gauge
transformations (IRGT), for the gauge \( F \). Let 'R' be the subspace of \( M^{4} \)
the translations along which leave \( \theta  \)'s of the IRGT's invariant.
We shall denote by \( \int d^{R}x \) the integration over this subspace of
\( M^{4} \). We may, for our convenience, write\footnote{%
This is a notation we introduce.
} \( \int d^{4}x\equiv \int d^{R}x\int d^{4-R}x \). 

\textbf{Examples of type -R gauges}:

\textbf{(1) The Coulomb family}:\( F=\nabla .A;\theta (x)=\theta (t) \);  with
either of the following satisfied:

(i) \( \theta (t) \) and \( \frac{d\theta }{dt} \) both \( \rightarrow 0 \)
as \( t\rightarrow \pm \infty  \), OR

(ii) \( \frac{d\theta }{dt} \) \( \rightarrow 0 \) and \( \theta (t) \)\( \rightarrow  \)
finite as \( t\rightarrow \pm \infty  \); if the boundary conditions on the
fields in the original path-integral (\ref{i}) are : \( \phi (\mathbf{x},t) \)
\( \rightarrow 0 \) as \( t\rightarrow \pm \infty  \) 

In this case, \( R=R^{3} \) {[}the 3-dimensional Euclidean space{]} and \( \int d^{R}x\equiv \int d^{3}\mathbf{x} \)
and \( \int d^{4-R}x\equiv \int dx_{0} \) .

\textbf{(2) An axial gauge}: \( F=A_{3};\theta (x)=\theta (x_{0},x_{1},x_{2}) \);
 with either of the following satisfied:

(i) \( \theta (x) \) and \( \frac{\partial \theta }{\partial x^{\mu }};\mu =0,1,2 \)
all \( \rightarrow 0 \) as \( t\rightarrow \pm \infty  \), OR

(ii) \( \frac{\partial \theta }{\partial x^{\mu }};\mu =0,1,2 \) \( \rightarrow 0 \)
and \( \theta (x) \)\( \rightarrow  \) finite as \( t\rightarrow \pm \infty  \);
if the boundary conditions on the fields in the original path-integral (\ref{i})
are : \( \phi (\mathbf{x},t) \) \( \rightarrow 0 \) as \( t\rightarrow \pm \infty  \)

In this case, \( R=R^{1} \)and \( \int d^{R}x\equiv \int dz \) and \( \int d^{4-R}x\equiv \int dx_{0}dx_{1}dx_{2} \).

\textbf{Theorem I}: The Minkowski space Lagrangian path-integral of (i) for
{}``type-R gauges{}'' leads to physically unacceptable results {[} such as
(\ref{rid1})--(\ref{rid3}) below{]}.

\textbf{Proof of Theorem I}: In the path-integral (\ref{i}), we perform, IRGT,
i.e. the transformations (\ref{brs}) on the integration variables. Noting that
(\ref{brs}) preserve the boundary conditions and the measure, we now see the
RGT-WT identity:\begin{equation}
\label{rwt}
<<\int d^{4}x\{J\delta A+K^{\dagger }\delta \psi +\delta \psi ^{\dagger }K+ghost\; source-terms\}>>=0
\end{equation}
where we have defined\footnote{%
We may be using the <\textcompwordmark{}<..>\textcompwordmark{}> notation \emph{with}
or \emph{without} ghost sources present; this has to be read in each specific
context.
}, for any \( O[\phi ] \),\[
<<O>>\equiv \int D\phi \, O[\phi ]exp\{iS_{eff}[A,c,\overline{c},\psi ]+source-terms\}\]
We now set the ghost sources to zero. A little simplification then allows one
to write:\begin{eqnarray}
0=<<\int d^{4}x\{J^{\alpha \mu }(x)D_{\mu }^{\alpha \beta }\theta ^{\beta }(x)\} &  & \nonumber \\
+\int d^{4}x\{K^{\dagger }(x)igT^{\alpha }\theta ^{\alpha }(x)\psi (x)-ig\psi (x)^{\dagger }T^{\alpha }\theta ^{\alpha }(x)K(x)>> & \label{1} 
\end{eqnarray}
We now express \( \int d^{4}x=\int d^{R}x\int d^{4-R}x \) and differentiate
with respect to \( \theta  \)\( ^{\beta } \)(x) (keeping in mind that the
translations in {}``R{}'' leave \( \theta  \)(x) invariant), we obtain:\begin{eqnarray}
0=<<\int d^{R}x\{-D_{\mu }^{\alpha \beta }(x)J^{\beta \mu }(x)\} &  & \nonumber \\
+\int d^{R}x\{K^{\dagger }(x)igT^{\alpha }\psi (x)-ig\psi (x)^{\dagger }T^{\alpha }(x)K(x)\}>> & \label{rwt0} 
\end{eqnarray}
We would like to show that these WT-identities lead to results that are physically
meaningless, and moreover are incompatible with those from the Lorentz gauge.
To show this, we choose a simpler special case: we specialize to the group SU(2),
let \( \psi  \) be a scalar in a complex doublet representation and and choose
the Coulomb gauge. Then {}``R{}'' is R\( ^{3} \)and \( T^{\alpha }=\frac{\sigma ^{\alpha }}{2} \).
Then, we have, \begin{eqnarray}
0=-\int d^{3}\mathbf{x}\{\frac{d}{dt}J_{0}^{\alpha }(\mathbf{x},t)-igJ_{\mu }^{\beta }(\mathbf{x},t)f^{\alpha \beta \gamma }\frac{\delta }{\delta J_{\mu }^{\gamma }(\mathbf{x},t)}\}\bullet W[J,K,K] &  & \nonumber \\
+\int d^{3}\mathbf{x}\left[ +gK_{i}^{*}(\mathbf{x},t)T_{ij}^{\alpha }\frac{\delta }{\delta K_{j}^{*}(\mathbf{x},t)}-gK_{j}(\mathbf{x},t)T_{ij}^{\alpha }\frac{\delta }{\delta K_{i}(\mathbf{x},t)}\right]  &  & \nonumber \\
\bullet W[J,K,K] & \label{rwt'} 
\end{eqnarray}
 Now differentiate with respect to \( K_{i}^{*}(\mathbf{z},t_{1}) \) and \( K_{l}(\mathbf{y},t') \)
with \( t'\neq t \) ; \( \mathbf{y}\neq \mathbf{z} \) and set \( J=K=K^{*}=0 \).
We then integrate over \( t_{1} \) from \( t-\delta  \) to \( t+\delta  \).
The point \( t' \) is outside this interval.We thus obtain\footnote{%
We keep \( \mathbf{y}\neq \mathbf{z} \) so that we do not have to face ill-defined
expressions. Moreover this procedure, in a general context, with the limit \( \delta \rightarrow 0 \)
taken, appears as taking the appropriate energy component to infinity \cite{prep}.
This will be used later. See comment (4) after Theorem IV.
},\begin{equation}
\label{rid1}
0=T_{ij}^{\alpha }\frac{\delta }{\delta K_{j}^{*}(\mathbf{z},t)\delta K_{l}(\mathbf{y},t')}W[J,K,K^{*}]\mid _{_{_{J=K=K*=0}}}
\end{equation}
Since, \( T^{\alpha }=\frac{\sigma ^{\alpha }}{2} \) are invertible matrices,
this would then mean that the propagator for the scalar field vanishes for \( \mathbf{z}\neq \mathbf{y} \)
for any \( t'\neq t \); evidently a conclusion incompatible with the corresponding
one for the Lorentz gauges\footnote{%
At this point, one may ask a justified question: would not the above conclusion
(\ref{rid1}) clash with one that would be obtained from the scalar propagator
in Lorentz gauges simply by a gauge transformation that takes one from the Lorentz
gauge to the Coulomb gauge?. We note that any such (careful) procedure starting
from the Lorentz gauge path-integral \emph{with} an \( \epsilon  \)-term will
\emph{not} land us with the \emph{naive (formal)} path-integral (\ref{i}) which
has lead to such results. For a careful treatment, see last of ref. \cite{jm'}).
}.

We could obtain further unacceptable consequences from (\ref{rwt'}). e.g.,
we could differentiate (\ref{rwt'}) with respect to \( J^{\alpha }_{0}(\mathbf{z},t' \))
and set all sources to zero. We are then lead to\begin{equation}
\label{rid2}
\frac{d}{dt}\delta (t-t')=0!
\end{equation}
which is even mathematically unsound. We could also try differentiation with
respect to \( J_{\nu }^{\delta }(w) \) and \( J_{\sigma }^{\rho }(v) \) and
then set all sources to zero. We will then arrive at a conclusion very similar
to that for a scalar propagator following (\ref{rid1}), viz. vanishing of the
gauge boson propagator \( \Delta _{\mu \nu }(v,w) \) for \( v\neq w \). 

\begin{equation}
\label{rid3}
\Delta _{\mu \nu }(v,w)=0;v\neq w
\end{equation}
 And so on.

\textbf{Comments on theorem I:}

The fact that the generating functional (\ref{i}) is ill-defined for Type-R
gauges is well-known. Nonetheless, it can lead to diverse absurd results by
mathematical manipulations \emph{along standard lines} has been exhibited above\footnote{%
A brief argument leading to the conclusion analogous to (\ref{rid1}), is also
found in ref. \cite{bz}.
}. The result above brings out the inevitable role that any corrective measure
in (\ref{i}) is likely to play. This is in fact born out by the results below!

\textbf{BRST WT-Identity and the \( \epsilon  \)-term}

Normally, even in Lorentz gauges, (\ref{i}) is understood to contain a suitable
\( \epsilon  \)-term to make the Green's functions well-defined; nevertheless
it is generally ignored in the formal manipulations and in the derivation of
WT-identities (and this is appropriate for the family of the Lorentz gauges
in the context of unbroken gauge theory \cite{j2001}). Suppose we modify (\ref{i})
as \begin{equation}
\label{ii}
W[J,K,\overline{K};\xi ,\overline{\xi }]=\int D\phi exp\{iS_{eff}[A,c,\overline{c},\psi ]+\varepsilon O[\phi ]+source-terms\}
\end{equation}
to include \emph{an} \( \epsilon  \)-term. We recognize that this \( \epsilon  \)-term
must, in particular, break the residual gauge invariance completely. In addition,
to keep the discussion general, we do not necessarily limit \( \epsilon  \)
to have dimension two in the following, nor do we restrict \( O \) to have
local nature\footnote{%
We do not however imply that \emph{any} such \( \epsilon  \)-term will necessarily
be appropriate to define a gauge theory compatible with the Lorentz gauges.
Existence ( and construction) of an \emph{}\( \epsilon  \)-term which will
serve \emph{this purpose} is already known \cite{jm2000,jm'} however. See also
comment (2) after the theorem IV. 
}. We note that the various prescriptions, say the Leibbrandt-Mandelstam for
the light-cone gauges and the CPV for axial gauges etc, can be understood\footnote{%
We however note some of the complications in the interpretation of \emph{double}
poles in CPV. See e.g. references \cite{bns,l}.
} as special cases of (\ref{ii}) {[}with rather complicated nonlocal \( O \){]}
and thus the following discussion includes these as special cases. .We then
we have the following result, (which we can well suspect from the theorem I):

\textbf{Theorem II}: The BRST WT-identity for (\ref{ii}) can receive a nontrivial
contribution from the \( \epsilon  \)-term for type -R gauges even as \( \epsilon \rightarrow 0 \).

Before we give the proof of the theorem II, we shall first prove two lemmas.

\textbf{Lemma 1: \( S_{eff} \)} satisfies Eq. \textbf{(\ref{inv})} below\textbf{.}

Invariance of \( S_{eff} \) under the IRGT of (\ref{brs}) implies that,\begin{eqnarray*}
\int d^{4}x\theta ^{\delta }\left[ D_{i}^{\delta }\frac{\delta }{\delta A_{i}}+gf^{\alpha \delta \beta }c^{\beta }\frac{\delta }{\delta c^{\alpha }}+gf^{\alpha \delta \beta }\overline{c}^{\beta }\frac{\delta }{\delta \overline{c}^{\alpha }}\right] S_{eff} &  & \\
=\int d^{4-R}x\int d^{R}x\theta ^{\delta }\left[ D_{i}^{\delta }\frac{\delta }{\delta A_{i}}+gf^{\alpha \delta \beta }c^{\beta }\frac{\delta }{\delta c^{\alpha }}+gf^{\alpha \delta \beta }\overline{c}^{\beta }\frac{\delta }{\delta \overline{c}^{\alpha }}\right] S_{eff} & = & 0
\end{eqnarray*}
Now, noting that the variation of \( \theta  \) belongs to space of functions
on \( M^{4}/R \),  we obtain,

\begin{equation}
\label{inv}
\int d^{R}x\left[ D_{i}^{\delta }\frac{\delta }{\delta A_{i}}+gf^{\alpha \delta \beta }c^{\beta }\frac{\delta }{\delta c^{\alpha }}+gf^{\alpha \delta \beta }\overline{c}^{\beta }\frac{\delta }{\delta \overline{c}^{\alpha }}\right] S_{eff}=0
\end{equation}

\textbf{Lemma 2:} Defining \[
Y^{\delta }[A,c,\overline{c},....]\equiv \frac{\delta S'_{eff}[A,c,\overline{c};.....]}{\delta c^{\delta }}\]
with \( S'_{eff} \) as the total effective action, \emph{including} the \( \epsilon  \)-term,
and \( V[\phi ] \) a Grassmann- odd functional, we have following results in
absence of ghost sources\footnote{%
For any operator \( O \), we define the BRS variation as \( \delta O\equiv (\delta _{BRS}O)\delta \Lambda  \).
}:\begin{eqnarray}
Y^{\delta }[-i\frac{\delta }{\delta J},-i\frac{\delta }{\delta \overline{\xi }},i\frac{\delta }{\delta \xi },.....]<<\{iJ_{i}(Dc)_{i}+.......\}+V[\phi ]>> &  & \nonumber \\
=<<\delta _{BRS}Y^{\delta }+i\frac{\delta }{\delta c^{\delta }}V[\phi ]-\left[ J_{i}D^{\delta }_{i}+.......\right] >> & \label{21} 
\end{eqnarray}
and\footnote{%
We note that henceforth the meaning of \( <<..>> \) is dependent on the specific
\( \epsilon  \)-term in (\ref{ii}).
}\begin{eqnarray}
<<\int d^{R}x\left\{ \delta _{BRS}Y^{\delta }+i\varepsilon \frac{\delta }{\delta c^{\delta }}\delta _{BRS}O[\phi ]\right\} -i\varepsilon \int d^{R}x(\Delta O)^{\delta }>>=0 & \label{22} 
\end{eqnarray}
Here \( \{iJ_{i}(Dc)_{i}+.......\} \) is the BRST variation of the source term
(\ref{src}), (that appears in the BRST WT-identity), with \( \delta \Lambda  \)
factored out to the right; and under IRGT, \( O\rightarrow O+\delta O\equiv O+\theta ^{\delta }(\Delta O)^{\delta } \). 

\textbf{Proof of Lemma 2}: Noting that \( Y^{\delta } \) is an odd Grassmannian,
we have,\begin{eqnarray*}
Y^{\delta }[-i\frac{\delta }{\delta J},-i\frac{\delta }{\delta \overline{\xi }},i\frac{\delta }{\delta \xi },.....]\{iJ_{i}(Dc)_{i}+.......\} &  & \\
=(\delta _{BRS}Y^{\delta })[-i\frac{\delta }{\delta J},-i\frac{\delta }{\delta \overline{\xi }},i\frac{\delta }{\delta \xi },.....] &  & \\
-\{iJ_{i}(Dc)_{i}+.......\}Y^{\delta }[-i\frac{\delta }{\delta J},-i\frac{\delta }{\delta \overline{\xi }},i\frac{\delta }{\delta \xi },.....] & 
\end{eqnarray*}
 Further for any Grassmann-odd \( U[\phi ] \), we have,\begin{eqnarray*}
Y^{\delta }[-i\frac{\delta }{\delta J},-i\frac{\delta }{\delta \overline{\xi }},i\frac{\delta }{\delta \xi },.....]U[\phi ]\exp \{iS_{eff}'+i\int d^{4}x[JA+....]\} &  & \\
=-U[\phi ]Y^{\delta }[A,c,\overline{c},....]\exp \{iS_{eff}'+i\int d^{4}x[JA+....]\} &  & \\
=-U[\phi ](-i\frac{\delta }{\delta c^{\delta }}+\overline{\xi }^{\delta })\exp \{iS_{eff}'+i\int d^{4}x[JA+....]\} & 
\end{eqnarray*}
We thus have, for a Grassmann-odd V{[}\( \phi  \){]}, \begin{eqnarray*}
Y^{\delta }[-i\frac{\delta }{\delta J},-i\frac{\delta }{\delta \overline{\xi }},i\frac{\delta }{\delta \xi },.....]<<\{iJ_{i}(Dc)_{i}+.......\}+V[\phi ]>> &  & \\
=<<\delta _{BRS}Y^{\delta }-\left[ \{iJ_{i}(Dc)_{i}+.......\}+V[\phi ]\right] (-i\frac{\delta }{\delta c^{\delta }}+\overline{\xi }^{\delta })>> &  & \\
=<<\delta _{BRS}Y^{\delta }+i\frac{\delta }{\delta c^{\delta }}\left[ \{iJ_{i}(Dc)_{i}+.......\}+V[\phi ]\right] +terms\, involving\, \overline{\xi }>> &  & \\
=<<\delta _{BRS}Y^{\delta }+i\frac{\delta }{\delta c^{\delta }}V[\phi ]-\left[ J_{i}D^{\delta }_{i}+.......\right] +terms\, involving\, \overline{\xi }>> &  & 
\end{eqnarray*}
This proves (\ref{21}). We now note\footnote{%
In the second step below, we have a commutator \( \left[ \frac{\delta }{\delta c^{\delta }},\delta _{BRS}\right]  \)
and not anticommutator; this is because of our convention regarding \( \delta  \)\( _{BRS} \)
.
}\begin{eqnarray*}
\delta _{BRS}Y^{\delta } & = & \delta _{BRS}\frac{\delta S'_{eff}[A,c,\overline{c}]}{\delta c^{\delta }}\\
 & = & \frac{\delta }{\delta c^{\delta }}\delta _{BRS}S_{eff}'[A,c,\overline{c}]-\left[ \frac{\delta }{\delta c^{\delta }},\delta _{BRS}\right] S_{eff}'[A,c,\overline{c}]\\
 & = & -i\varepsilon \frac{\delta }{\delta c^{\delta }}\delta _{BRS}O[\phi ]-\left[ D_{i}^{\delta }\frac{\delta }{\delta A_{i}}-gf^{\alpha \delta \beta }c^{\beta }\frac{\delta }{\delta c^{\alpha }}\right] S_{eff}'[A,c,\overline{c}]
\end{eqnarray*}
We now employ the invariance of \( S_{eff} \) under IRGT as expressed by (\ref{inv})
to simplify the last term as \begin{eqnarray*}
\int d^{R}x\left[ D_{i}^{\delta }\frac{\delta }{\delta A_{i}}-gf^{\alpha \delta \beta }c^{\beta }\frac{\delta }{\delta c^{\alpha }}\right] S'_{eff} &  & \\
=\int d^{R}x\left[ D_{i}^{\delta }\frac{\delta }{\delta A_{i}}+gf^{\alpha \delta \beta }c^{\beta }\frac{\delta }{\delta c^{\alpha }}+gf^{\alpha \delta \beta }\overline{c}^{\beta }\frac{\delta }{\delta \overline{c}^{\alpha }}\right] S'_{eff} &  & \\
-\int d^{R}x\left[ 2gf^{\alpha \delta \beta }c^{\beta }\frac{\delta }{\delta c^{\alpha }}+gf^{\alpha \delta \beta }\overline{c}^{\beta }\frac{\delta }{\delta \overline{c}^{\alpha }}\right] S'_{eff} &  & \\
=\int d^{R}x\left[ D_{i}^{\delta }\frac{\delta }{\delta A_{i}}+gf^{\alpha \delta \beta }c^{\beta }\frac{\delta }{\delta c^{\alpha }}+gf^{\alpha \delta \beta }\overline{c}^{\beta }\frac{\delta }{\delta \overline{c}^{\alpha }}\right] (-i\varepsilon O) &  & \\
-\int d^{R}x\left[ 2gf^{\alpha \delta \beta }c^{\beta }\frac{\delta }{\delta c^{\alpha }}+gf^{\alpha \delta \beta }\overline{c}^{\beta }\frac{\delta }{\delta \overline{c}^{\alpha }}\right] S'_{eff} &  & \\
=-i\varepsilon \int d^{R}x(\Delta O)^{\delta }+terms\, whose\, expectation\, vanishes\, at\, \xi =\overline{\xi }=0 & 
\end{eqnarray*}
We thus have, in absence of ghost sources,\begin{eqnarray}
<<\int d^{R}x\left\{ \delta _{BRS}Y^{\delta }+i\varepsilon \frac{\delta }{\delta c^{\delta }}\delta _{BRS}O[\phi ]\right\} >> & = & i\varepsilon \int d^{R}x<<(\Delta O)^{\delta }>>\label{y} 
\end{eqnarray}

\textbf{Proof of theorem II:}

We now apply the procedure used in the proof of theorem I, that lead to (\ref{rwt'}),
to the generating functional of (\ref{ii}) that incorporates an \( \epsilon  \)-term.
Evidently, the derivation of the WT-identity for the IRGT from eq. (\ref{rwt})
to (\ref{rwt'}) now involves the variation of this term under IRGT. Then, (\ref{rwt'})is
now replaced by\begin{eqnarray}
0=-\int d^{3}\mathbf{x}\{\frac{d}{dt}J_{0}^{\alpha }(\mathbf{x},t)-igJ_{\mu }^{\beta }(\mathbf{x},t)f^{\alpha \beta \gamma }\frac{\delta }{\delta J_{\mu }^{\gamma }(\mathbf{x},t)}\}\bullet W[J,K,K] &  & \nonumber \\
+\int d^{3}\mathbf{x}\left[ gK_{i}^{*}(\mathbf{x},t)T_{ij}^{\alpha }\frac{\delta }{\delta K_{j}^{*}(\mathbf{x},t)}-gK_{j}(\mathbf{x},t)T_{ij}^{\alpha }\frac{\delta }{\delta K_{i}(\mathbf{x},t)}\right]  &  & \nonumber \\
\bullet W[J,K,K] &  & \nonumber \label{rwt1} \\
-i\varepsilon <<(\Delta O)^{\alpha }(\mathbf{x},t)>> &  & \label{BRS} 
\end{eqnarray}
 Evidently, to avoid physically unacceptable results such as (\ref{rid1})-(\ref{rid3}),
the \( \epsilon  \)-term must be contributing to the IRGT WT- identities in
a non-trivial way. 

We now wish to correlate the result above (IRGT WT-identity) with what we normally
understand by the BRST WT-identity; taking care of the \( \epsilon  \)-term
in both, however. To do this, we write down the BRST WT-identity for the generating
functional (\ref{ii}) :viz\begin{eqnarray}
<<\int d^{4}x\left\{ J_{\mu }^{\alpha }(x)D_{\mu }^{\alpha \beta }c^{\beta }(x)-\overline{\xi ^{\alpha }(x)}\frac{1}{2}gf^{\alpha \beta \gamma }c^{\beta }c^{\gamma }(x)-\frac{1}{\lambda }\xi ^{\alpha }F^{\alpha }(x)\right\}  &  & \nonumber \\
+\int d^{4}x\left\{ K^{\dagger }(x)igT^{\alpha }c^{\alpha }(x)\psi (x)+....\right\} -i\varepsilon \delta _{BRS}O>> & = & 0\label{BRST} 
\end{eqnarray}
We shall find it convenient to use the De-Witt summation-integration convention,
say as in ref.\cite{jl} and write:\begin{eqnarray*}
0=<<\left\{ J_{i}D_{i}^{\alpha }c^{\alpha }-\overline{\xi ^{\alpha }}\frac{1}{2}gf^{\alpha \beta \gamma }c^{\beta }c^{\gamma }-\frac{1}{\lambda }\xi ^{\alpha }F^{\alpha }+igK_{i}^{\dagger }T^{\alpha }_{ij}\psi _{j}c^{\alpha }-ig\psi _{i}^{\dagger }T^{\alpha }_{ij}K_{j}c^{\alpha }\right\}  &  & \\
-i\varepsilon \delta _{BRS}O>> & 
\end{eqnarray*}
We now act on it by \( -Y^{\delta }[-i\frac{\delta }{\delta J},-i\frac{\delta }{\delta \overline{\xi }},i\frac{\delta }{\delta \xi },.....] \)
 and set \( \xi =0=\overline{\xi } \) afterwords. Using (\ref{21}), and identifying
\( V[\phi ]=\varepsilon \delta _{BRS}O \) , we obtain\[
0=<<\left\{ J_{i}D_{i}^{\alpha }+igK_{i}^{*}T^{\alpha }_{ij}\psi _{j}-ig\psi _{i}^{*}T^{\alpha }_{ij}K_{j}\right\} -i\varepsilon \frac{\delta }{\delta c^{\alpha }}\delta _{BRS}O-\delta _{BRS}Y^{\alpha }>>\]
We now integrate the above equation over the subspace {}``\( R \){}'' of
\( M^{4} \)and employ (\ref{22}) to arrive at the following from the BRST
WT-identity:\begin{equation}
\label{rgi2}
0=<<\int d^{R}x\left\{ J_{i}D_{i}^{\alpha }+igK_{i}^{*}T^{\alpha }_{ij}\psi _{j}-ig\psi _{i}^{*}T^{\alpha }_{ij}K_{j}\right\} -i\varepsilon \int d^{R}x(\Delta O)^{\alpha }>>\mid _{_{\xi =0=\overline{\xi }}}
\end{equation}
Thus, the IRGT WT-identity is a particular consequence of the BRST WT-identity
(and (\ref{BRS}) is its special application in connection with the Coulomb
gauge). We have already seen from the Theorem I that the correct IRGT WT-identity
must receive a nontrivial contribution from the \( \epsilon  \)-term even as
\( \epsilon \rightarrow 0 \), without which it will lead to several incorrect
conclusions. Now, if it could have been possible to drop the \( \epsilon  \)-term
from (\ref{BRST}), (as also from the equations of motion), then the correct
\( \epsilon  \)-term as in (\ref{BRS}) would not have appeared in (\ref{rgi2}). 

\textbf{Conditions Fulfilled by \( \epsilon  \)-term}

The BRST WT-identity for (\ref{ii}) also spells out from (\ref{rgi2}) further
algebraic conditions that are implied for the expectation value of the \( \epsilon  \)-term
in (\ref{ii}) alone. 

\textbf{Theorem III: \( <<\int d^{R}x(\Delta O)^{\alpha }>> \)} appearing in
(\ref{rgi2}) must satisfy\[
\int d^{R}x\int d^{R}y\left\{ X^{\alpha }<<(\Delta O)^{\beta }>>-X^{\beta }<<(\Delta O)^{\alpha }>>\right\} =gf^{\alpha \beta \gamma }\delta ^{(4-R)}(x-y)<<\int d^{R}x(\Delta O)^{\gamma }>>\]
where \( X \) has been defined below in (\ref{rgi3}).

\textbf{Proof of theorem III:}

Using relations such as \[
<<\int d^{R}x\left\{ J_{i}D_{i}^{\alpha }\right\} >>=\int d^{R}x\left\{ J_{i}D_{i}^{\alpha }[-i\frac{\delta }{\delta J}]\right\} W[J,K,K^{*}]\]
we express (\ref{rgi2}) as\footnote{%
We note that the definition of X below contains sources only of gauge and matter
fields and not of ghost fields.
} \begin{equation}
\label{rgi3}
0=\int d^{R}xX^{\alpha }W[J,K,K^{*}]\equiv \int d^{R}x\left\{ J_{i}D_{i}^{\alpha }[-i\frac{\delta }{\delta J}]+......\right\} W[J,K,K^{*}]=<<i\varepsilon \int d^{R}x(\Delta O)^{\alpha }>>
\end{equation}
We note the following gauge algebra, expressed successively with summation-integration
convention and without it and with obvious notations,\begin{equation}
\label{alg}
[X^{\alpha },X^{\beta }]=gf^{\alpha \beta \gamma }X^{\gamma }=gf^{\alpha \beta \gamma }\int d^{4}z\delta ^{(4)}(x-y)\delta ^{(4)}(x-z)X^{\gamma }=gf^{\alpha \beta \gamma }\delta ^{(4)}(x-y)X^{\gamma }
\end{equation}
We now operate on (\ref{rgi3}) by \( \int d^{R}yX^{\beta } \) and take the
difference as \( (\alpha \leftrightarrow \beta ) \). We then have,\[
\int d^{R}x\int d^{R}y\left[ X^{\alpha },X^{\beta }\right] W[J,K,K^{*}]=i\varepsilon \int d^{R}x\int d^{R}y\left\{ X^{\alpha }<<(\Delta O)^{\beta }>>-X^{\beta }<<(\Delta O)^{\alpha }>>\right\} \]
We now employ (\ref{alg}) (in the expanded form), and carry out the \( \int d^{R}y \)
integration to find,\begin{eqnarray*}
gf^{\alpha \beta \gamma }\delta ^{(4-R)}(x-y)\int d^{R}xX^{\gamma }W[J,K,K^{*}] & = & gf^{\alpha \beta \gamma }\delta ^{(4-R)}(x-y)<<i\varepsilon \int d^{R}x(\Delta O)^{\gamma }>>\\
 & = & i\varepsilon \int d^{R}x\int d^{R}y\left\{ X^{\alpha }<<(\Delta O)^{\beta }>>-X^{\beta }<<(\Delta O)^{\alpha }>>\right\} 
\end{eqnarray*}
This leads to the condition\[
\int d^{R}x\int d^{R}y\left\{ X^{\alpha }<<(\Delta O)^{\beta }>>-X^{\beta }<<(\Delta O)^{\alpha }>>\right\} =gf^{\alpha \beta \gamma }\delta ^{(4-R)}(x-y)<<\int d^{R}x(\Delta O)^{\gamma }>>\]

We note that the above condition expresses simply a relation involving the matrix
elements of \( <<(\Delta O)^{\alpha }>> \)\footnote{%
We note, in particular, our earlier remark that the meaning of \( <<...>> \)is
itself dependent on the \( \varepsilon O \)-term.
}. We do expect that, at least for the correct \( \epsilon  \)-term, the above
condition is compatible in form with existence of a renormalization procedure.

\textbf{Additional Constraints implied by WT-Identity:}

We now wish to demonstrate that there are additional constraints that are implied
by the IRGT WT-identity {[} which are, of course, implicit in the \( \epsilon  \)-dependent
BRST WT-identity (\ref{BRST}){]} which have to be taken care of, and that the
form of these relates to the specific form of the \( \epsilon  \)-term. These
identities will, in particular, tell us how the propagator will be treated near
its singularity\footnote{%
These relations are expected to get modified by renormalization, in asmuchas
the \( \epsilon  \)-term will be modified by it, should renormalization for
the theory be possible with the \( \epsilon  \)-term at hand. See comment (5)
after theorem IV.
}. To illustrate this, suppose we were to assume, as is often done (explicitly
or in effect), that the \( \epsilon  \)-term in (\ref{ii}) is a general local
quadratic term \footnote{%
If we assume that the infinitesimal parameter \( \epsilon  \) is of dimension
two, then locality, zero ghost number and global gauge invariance would restrict
it to this quadratic form in any case. We do not, however, necessarily imply
that this \( \epsilon  \)-term is the correct one for the generating functional
(\ref{ii}). 
} :

\begin{equation}
\label{fe}
\varepsilon O[\phi ]=\varepsilon [\frac{1}{2}a_{ij}A_{i}A_{j}+b\overline{c}^{\alpha }c^{\alpha }]
\end{equation}
Then we have the following result:

\textbf{Theorem IV}: The local quadratic \( \epsilon  \)-term (\ref{fe}) implies
additional constraints on Green's functions that are derivable from (\ref{rgi2})
for the type-R gauges.

\textbf{Proof of Theorem IV:}

We have,  under IRGT of (\ref{brs}), \begin{eqnarray*}
\varepsilon (\Delta O)^{\alpha }= & \varepsilon a_{ij}A_{i}\partial _{j}^{\alpha }
\end{eqnarray*}
Now we shall illustrate the proof for the Coulomb gauge. The \( \epsilon  \)-term
(\ref{fe}) then leads to a term in the IRGT WT-identity of (\ref{BRS}) depending
on \begin{equation}
\label{fe1}
-i\int d^{R}x\varepsilon (\Delta O)^{\alpha }=-i\int d^{3}\mathbf{x}\varepsilon (\Delta O)^{\alpha }=i\varepsilon \int d^{3}\mathbf{x}a_{\nu 0}\frac{\partial }{\partial t}A^{\alpha }_{\nu }(\mathbf{x},t)
\end{equation}
Now,\begin{equation}
\label{fe2}
<<i\varepsilon \int d^{3}\mathbf{x}a_{\nu 0}\frac{\partial }{\partial t}A^{\alpha }_{\nu }(\mathbf{x},t)>>=\varepsilon \int d^{3}\mathbf{x}a_{\nu 0}\frac{\partial }{\partial t}\frac{\delta W}{\delta J_{\nu }^{\alpha }(\mathbf{x},t)}
\end{equation}
Suppose we now differentiate (\ref{BRS}) with respect to \( J_{\sigma }^{\alpha }(\mathbf{y},t') \)
and set all sources to zero, we find ( \( \alpha  \) not summed):\begin{equation}
\label{t}
-\frac{d}{dt}\delta (t-t')g_{\sigma 0}+\varepsilon \int d^{3}\mathbf{x}a_{\nu 0}\frac{\partial }{\partial t}\frac{\delta ^{2}W}{\delta J_{\nu }^{\alpha }(\mathbf{x},t)\delta J_{\sigma }^{\alpha }(\mathbf{y},t')}\mid _{_{_{0=J=...}}}=0
\end{equation}
This is a {}``corrected{}'' version of (\ref{rid2}); and similar {}``corrected{}''
versions will follow for (\ref{rid1}) and (\ref{rid3}). 

Evidently, (\ref{t}) this implies a constraint on the unrenormalized propagator
\( \Delta _{\nu \sigma }(p) \) to all orders:\[
\varepsilon p_{0}a_{\nu 0}i\Delta _{\nu \sigma }(\mathbf{p}=0,p_{0})=-g_{\sigma 0}p_{0}\]
Thus, for \( p_{0}\neq 0 \),  we have the constraint on \( \Delta _{\nu \sigma }(\mathbf{p}=0,p_{0}) \)
and \( a_{\nu 0} \) that \[
\varepsilon a_{\nu 0}\Delta _{\nu \sigma }(\mathbf{p}=0,p_{0})=ig_{\sigma 0}\]
Such a constraint is satisfied \emph{by the unrenormalized propagator to all
orders.} There is no parallel for such a constraint in Lorentz gauges.

We note that the above constraint is trivially true in the lowest order, if
the propagators are made well-defined by \emph{such an \( \epsilon  \)-term.}
For example, if we let \( a_{\mu \nu } \) diagonal, then for \( \sigma =0 \),
 we have\begin{eqnarray}
\varepsilon a_{\nu 0}\Delta _{\nu 0}(\mathbf{p}=0,p_{0})=\varepsilon a_{00}\Delta _{00}(\mathbf{p}=0,p_{0})=\frac{\varepsilon a_{00}}{|\mathbf{p}|^{2}-i\varepsilon a_{00}}\mid _{_{_{_{|\mathbf{p}|=0}}}} & =ig_{00} & \nonumber 
\end{eqnarray}
 and for \( \sigma =i \),  we have \[
\varepsilon a_{j0}\Delta _{ji}(\mathbf{p}=0,p_{0})=0=ig_{i0}\]

More relations follow along these lines; for example, those that will {}``correct{}''
the absurd relations (\ref{rid1}) and (\ref{rid3}) etc. As seen in the above
example, these relations refer to Green's functions with (some) momenta in appropriate
{}``singular{}'' subspaces ( e.g. \( \eta .k=0 \) for axial gauges; \( |\mathbf{k}|=0 \)
for the Coulomb gauge etc.)

To understand the meaning of these additional relations, we note the following:
A type-R gauge is specified by \emph{two} things: (i) The BRST-invariant effective
action \( S_{eff} \) ,(ii) The way of interpreting the singularities in propagators:
viz. the {}``prescription{}'' or the \( \epsilon  \)-term \( \varepsilon O[\phi ] \),
the symmetry breaking term, as formulated in the path-integral (\ref{ii}).
We know of several examples which show that the latter (i.e. prescription) crucially
affects the divergence structure, gauge-invariance and the renormalization properties
\cite{bns,l}. The \( \epsilon  \)-dependent BRST WT-identity (\ref{BRST})
contains \emph{both} the consequence of the BRST invariance of \( S_{eff} \)
\emph{and} the effect of a specific \( \epsilon  \)-term \( \varepsilon O[\phi ] \).
The extra relations obtained above refer to the latter. The renormalization
of the theory (at least for the correct \( \epsilon  \)-term), must deal with
both of these. See comment (5) below.

\textbf{Comments on theorems II and IV:}

(1) Any \( \epsilon  \)-term (\ref{ii}) that breaks the IRGT does in fact
determine some generating functional \( W \) ( as a mathematical entity) for
which the {}``Green's functions{}'' will satisfy additional relations of the
sort (\ref{t}) . This applies to various {}``prescriptions{}'' for dealing
with such gauges, if they can be put in the form (\ref{ii}) and are representative
of that {}``prescription{}''. It is far from obvious, of course, that such
a generating functional will represent a physical theory compatible with the
Lorentz gauges for \emph{every} such choice.

(2) A way to determine correct \( \epsilon  \)-term (i.e. one compatible with
the Lorentz gauges) has been presented in \cite{jm2000} based on earlier works
\cite{jm}. It has been variously exploited in connection with the non-covariant
gauges \cite{jm'}. 

(3)That the \( \epsilon  \)-term can contribute to the WT-identity and that
not any simple \( \epsilon  \)-term can do the job has been recently demonstrated
in connection with the Doust gauge \cite{j2001}. 

(4)In passing, we note that the \( \epsilon  \)-term contributes to \emph{even
the tree WT-identity} with the above \( \epsilon  \)-term as seen trivially
in the example below: We first note \begin{eqnarray}
\frac{1}{k^{2}-m^{2}+i\varepsilon }-\frac{1}{(k+q)^{2}-m^{2}+i\varepsilon } &  & \label{lhs} \\
=\frac{q.(q+2k)}{k^{2}-m^{2}+i\varepsilon }\frac{1}{(k+q)^{2}-m^{2}+i\varepsilon } &  & \nonumber \\
=\frac{-\mathbf{q}.(\mathbf{q}+2\mathbf{k})}{k^{2}-m^{2}+i\varepsilon }\frac{1}{(k+q)^{2}-m^{2}+i\varepsilon }+\frac{q_{0}.(q+2k)_{0}}{k^{2}-m^{2}+i\varepsilon }\frac{1}{(k+q)^{2}-m^{2}+i\varepsilon } &  & \nonumber 
\end{eqnarray}
and that the expression (\ref{lhs}), at \textbf{q=0,} equals \textbf{} the
following at \textbf{q=0.} \begin{equation}
\label{e}
\frac{-\mathbf{q}.(\mathbf{q}+2\mathbf{k})}{k^{2}-m^{2}+i\varepsilon }\frac{1}{(k+q)^{2}-m^{2}+i\varepsilon }+\varepsilon q_{0}\frac{i}{|\mathbf{q}|^{2}+i\varepsilon }\frac{(q+2k)_{0}}{k^{2}-m^{2}+i\varepsilon }\frac{1}{(k+q)^{2}-m^{2}+i\varepsilon }
\end{equation}
The expression (\ref{lhs}) at \textbf{q=0} is arising from the second and the
third term in (\ref{rgi2}) while the last term in (\ref{e}) is arising from
the last term in (\ref{rgi2}); {[}see Eq. (\ref{fe2}){]}. If, on the other
hand, we put \( \mathbf{q}=0 \) in (\ref{lhs}) and take the limit \( q_{0}\rightarrow \infty  \),
what is left of (\ref{lhs}) is the propagator \( \frac{1}{k^{2}-m^{2}+i\varepsilon } \)
while under these limits, (\ref{e}) reduces to the same expression. Here, we
note that the first term is the momentum space (tree) expression for the right
hand side of (\ref{rid1}) while \emph{second term has now arisen from the \( \epsilon  \)}-term
in (\ref{rgi2}) as \( q_{0}\rightarrow \infty  \) {[} See footnote earlier
before Eq.(\ref{rid1}){]}. Without this \( \epsilon  \)-term, we would be
left with the absurd relation (\ref{rid1}).

(5) We note that the WT-identity (\ref{BRST}), with additional sources added
for the BRST variations, can be reexpressed as the \emph{modified} Zinn-Justin
equation for the total effective action \( S'_{eff} \) , with the \( \epsilon  \)-term
treated as just another term (a symmetry breaking term) in it with additional
parameter \( \epsilon  \):

\begin{equation}
\label{zj}
\Gamma *\Gamma =i\varepsilon <\delta _{BRS}O>
\end{equation}
Here the last term has been expressed as a functional of field expectation values
and BRST-variation sources. A method is to be developed along the standard lines
to deal with the renormalization\footnote{%
The \( \epsilon  \)-term could be renormalized, and we may also allow rescaling
of \( \varepsilon  \).
} of the above order by order \cite{prep}. Whether BRST will actually be maintained
for any {}``prescription{}'' (or equivalently any choice of \( O[\phi ] \))
depends on (the possibility of carrying out renormalization in the first place
and on) compatibility of renormalization of \( \delta _{BRS}O \) and renormalization
of fields with (\ref{zj}); and it is far from obvious that it will work generally.
In this connection, we draw attention to the formalism developed in \cite{jm2000}
and further pursued in \cite{jm'}; as it starts off from a BRST-invariant formalism
of the Lorentz gauges and does not {}``impose from \emph{outside}{}'' a prescription
to deal with these gauges (and therefore is likely to be BRST-consistent). We
wish to pursue this further elsewhere.

(6) There may be ways of defining the type-R gauges (by breaking IRGT) that
cannot be expressed by an \( \epsilon  \)-term in the path-integral. In these
cases also, we would expect a series of relations of the kind written down in
the proof of the Theorem IV.

\textbf{A Comment On Wick Rotation:}

We now discuss the possibility that the generating functional of the form (\ref{ii}),
with correct \( \epsilon  \)-term, can be obtained through a Wick rotation
from an Euclidean space generating functional that need not contain any further
\( \epsilon  \)-term. We restrict ourselves to those gauges for which the Wick-rotated
gauge function \( F^{E} \) is either {}``real{}'' or purely imaginary\footnote{%
This keeps \( F^{2} \) real. An imaginary factor in the ghost action \( S_{gh} \)
is of no consequence.
}. This includes gauges such as the Coulomb gauge, the spatial axial gauge {[}\( \eta =(0,\overrightarrow{\eta })] \)
and the temporal gauge \( A_{0}=0 \) ; but does not include the light-cone
gauge or a \emph{general variety} axial gauge with \( F=\eta .A \) (with \( \eta  \)\( _{0} \)
and \( \overrightarrow{\eta } \) both non-vanishing). We now consider if the
gauge F is of type-R in the Euclidean space. If so, we shall call it type-R\( ^{E} \).
We normally imagine the Euclidean space path-integral as \begin{equation}
\label{eu}
W[J,K,\overline{K},\xi ,\overline{\xi }]=\int D\phi exp\{-S^{E}_{eff}[A,c,\overline{c},\psi ]+source-terms\}
\end{equation}
without any further \( \epsilon  \)-terms to make the path-integral well-defined.
We ask whether we could have obtained the correct path-integral as in (\ref{ii})
for these gauges that includes an appropriate \( \epsilon  \)-term by Wick
rotation starting from such a Euclidean path-integral as in (\ref{eu}). We
see that for gauges which are also type-R\( ^{E} \), the Euclidean space path-integral
of (\ref{eu}) also leads to the similar problems, brought out in theorem I,
as those for the generating functional (\ref{i}) in the context of the Minkowski
space . Thus, if (in the context of such gauges), the Wick rotation is a meaningful
process we do not expect it to lead from a meaningful generating functional
(\ref{ii}) in Minkowski space into (\ref{eu}) which leads to incorrect relations
as (\ref{rid1}) after conversion to the Euclidean space. This argues against
the possibility that for these gauges, simultaneously type-R and type-R\( ^{E} \),
the correct Minkowski space generating functional (\ref{ii}) could be obtained
from (\ref{eu}) by a Wick rotation. We note that the Coulomb, the pure axial
and the (pure) temporal gauges are examples of such gauges.

This argument speaks in favor of the procedure adopted in \cite{jm2000} of
\emph{constructing} the \emph{Minkowski space} path-integral \emph{}(at least
for such gauges) by relating these to the \emph{Minkowski space path-integral}
for Lorentz gauges\emph{.} Moreover, the above results indicate that the general
formulation starting from (\ref{ii}) may be advantageous in formulating these
gauges.

\textbf{ACKNOWLEDGMENT}

I would like to acknowledge support from Department of Science and Technology,
India via grant for the project No. DST/PHY/19990170.


\begin{thebibliography}{10}
\bibitem{bns}A. Bassetto, G. Nardelli, and R. Soldati, Yang-Mills Theories in Algebraic Noncovariant
Gauges (World Scientific, Singapore, 1991).
\bibitem{l}G. Leibbrandt, Noncovariant Gauges (World Scientific, Singapore, 1994).
\bibitem{1}See  references 4-8 for some recent works; also see the references in \cite{bns,l}.
\bibitem{jm}S. D. Joglekar, and B. P. Mandal, Phys. Rev. D51, 1919-1928 (1995); R. S. Bandhu,
and S. D. Joglekar, J. Phys. A31, 4217-4224 (1998).S. D. Joglekar Ind.J.Phys.
73B,137(1999)
\bibitem{bz}L. Baulieu, and D. Zwanziger, Nucl. Phys. B548, 527-562 (1999).
\bibitem{jm'}S. D. Joglekar, and A. Misra, Mod. Phys. Lett. A14, 2083 (1999); \textbf{\emph{ibid}}
A15, 541-546 (2000); Int. J. Mod. Phys.A 16, 3731 (2001); S. D. Joglekar, Mod.
Phys. Lett. A15, 245-252 (2000); Int. J. Mod. Phys.A 16, 5043 (2001); S. D.
Joglekar, and B. P. Mandal, Int. J. Mod. Phys.A 17, 1279 (2002).
\bibitem{jm2000}S. D. Joglekar, and A. Misra, Int. J. Mod. Phys.A 15, 1453 (2000); Erratum \emph{ibid}A15,3899(2000);
S. D. Joglekar, and A. Misra, J. Math. Phys. 41, 1755-1767 (2000).
\bibitem{j2001}S. D. Joglekar, Euro.Phys. Journal-direct C12 3, 1-18 (2001); and hep-th/0201011.
\bibitem{iz}C. Itzykson and J.B.Zuber, \emph{Quantum field theory} ( McGraw Hill, New York,1985)
\bibitem{egbz}See e.g. \cite{bz} and references therein.
\bibitem{gsw}See e.g. M.Green, J.Schwartz, E. Witten \emph{Superstring theory ,} (Cambridge
university press, Cambridge, 1987)
\bibitem{jl}See e.g. S.D.Joglekar and B.W.Lee, Ann. Phys. 97,160(1976) and B.W.Lee Phys.Rev.D9,
933 (1974)
\bibitem{prep}S.D.Joglekar (in preparation).\end{thebibliography}
\end{document}